\documentclass[draft,jgrga]{agutex}
\usepackage{graphicx}
\usepackage{epstopdf}
\usepackage{amsmath,amssymb,amsfonts}
\usepackage[usenames]{color}
\authorrunninghead{Sasunov ET AL.}

\titlerunninghead{scaling of a thin current sheet}

\authoraddr{Yu. L. Sasunov, Space Research Institute, Austrian Academy of
Sciences, Graz, Austria. (jury.sasunov@oeaw.ac.at)}

\authoraddr{M. L. Khodachenko, Space Research Institute, Austrian Academy of
Sciences, Graz, Austria. (maxim.khodachenko@oeaw.ac.at)}

\authoraddr{I. I. Alexeev, Institute of Nuclear Physic, Moscow,
Russia. (alexeev@dec1.sinp.msu.ru)}

\authoraddr{E. S. Belenkaya, Institute of Nuclear Physic, Moscow,
Russia. (elena@dec1.sinp.msu.ru)}

\authoraddr{V. S. Semenov, Saint-Petersburg State University, Saint-Petersburg, Russia (sem@geo.phys.spbu.ru)}

\authoraddr{I. V. Kubyshkin, Saint-Petersburg State University, Saint-Petersburg, Russia (kubh@geo.phys.spbu.ru)}

\authoraddr{O. V. Mingalev, Polar Geophysical Institute, Kola Scientific Center, Russian Academy of
Sciences, Apatity, Russia (mingalev$_{-}$o@pgia.ru)}

\begin{document}

%
%
\setkeys{Gin}{draft=false}
%
%

\title{Investigation of scaling properties of a thin current sheet by means of particle trajectories study}
%
%
\authors{Yu. L. Sasunov, \altaffilmark{1} M. L. Khodachenko, \altaffilmark{1},\altaffilmark{2}
I. I. Alexeev, \altaffilmark{3} E. S. Belenkaya, \altaffilmark{3} V. S. Semenov, \altaffilmark{4} I. V. Kubyshkin, \altaffilmark{4}, O. V. Mingalev,\altaffilmark{5}}

\altaffiltext{1}{Space Research Institute, Austrian Academy of Sciences, Graz, Austria.}
\altaffiltext{2}{Skobeltsyn Institute of Nuclear Physics, Moscow State University, Moscow, Russia.}
\altaffiltext{3}{Institute of Nuclear Physic, Moscow, Russia.}
\altaffiltext{4}{Saint-Petersburg State University, Saint-Petersburg, Russia.}
\altaffiltext{5}{Polar Geophysical Institute, Kola Scientific Center, RAS, Apatity, Russia.}
%
%

\begin{abstract}
A thin current sheet (TCS), with the width of an order of thermal proton gyroradius, appears a fundamental physical object which plays an important role in structuring of major magnetospheric current systems (magnetotail, magnetodisk, etc.). The TCSs are nowadays under extensive study by means of space missions and theoretical models. We consider a simple model of the TCS separating two half-spaces occupied by a homogenous magnetic field of opposite sign tangential to the TCS; a small normal component of the magnetic field is prescribed. An analytical solution for the electric current and plasma density in the close vicinity of the TCS has been obtained and compared with numerical simulation. The number density and the electric current profiles have two maxima each. The characteristic spatial scale $z_S$ of the maxima location was investigated as a function of initial pitch-angle of an incoming charge particle. The effect of the thermal dispersion of the incoming proton beam have been taken into consideration.
\end{abstract}

%
%

\begin{article}

\section{Introduction}\label{sec:intro}
%
%
One of the challenges of modern space physics is understanding of the formation and fine structure of current sheets (CSs). The dynamical evolution of CSs and related energetic processes are responsible for substorms, solar flares, and other space plasma phenomena. In fact, CSs appear an accumulator of magnetic energy, thus the processes of the CSs decay consist in the transformation of magnetic energy to kinetic and thermal energy of plasma. The experimental information regarding CSs is nowadays provided by space missions, specifically those investigating the Earth's magnetosphere and solar wind (e.g., CLUSTER, GEOTAIL, THEMIS etc.).

The observation of evolution of a CS in Earth's magnetail shows that during a substorm the thickness of the CS could have a few hundred kilometers \citep[]{Mitchell1990, Artemyev2010b, Nakamura2008, Sergeev1990}, which is of the order of magnitude of a few Larmor radii of proton. Such thin current sheets (TCSs) have been also observed in Mercury \citep[]{Whang1997}, solar wind \citep[]{Gosling2008}, Jovian magnetodisk \citep[]{Alexeev2005} and they are expected to play an important role in magnetospheres of exoplanets \citep[]{Khodachenko2012}. Numerous in situ observations performed by spacecraft, report the presence of a certain spatial structuring of TCSs, e.g. the presence of anisotropic plasma flows and a multi-scale character of the electric current distribution nested in the TCSs \citep[]{Runov2006, Runov2003}. The expected reconnection processes and instabilities occurring in the TCSs are believed to be a driver for all the above mentioned large-scale astrophysical energetic and dynamical phenomena. This justifies an actuality of a further investigation of TCSs with a thickness of a few thermal proton gyroradii. Analytical methods in that respect appear a powerful tool which connects the studied natural phenomenon of a TCS with the backgrounds of fundamental plasma physics. As it will be shown in this paper, the estimation of thickness of a TCS as a function of the ratio of the incoming flow velocity to the thermal speed can be performed with the trajectory method.

Theoretical studies of a TCS in application to various space plasma problems are related with the ideas of Syrovatskii \citep[]{Syrovatskii1971} regarding the MHD formation and dynamics of a TCS near the magnetic field X-line. However, the MHD description is inapplicable at small scales comparable to the thermal Larmor radius of a proton. In that respect, the plasma kinetic approach is more correct here and may be expected to give physically reliable results regarding the processes in the TCSs.

An attempt of a direct numeric simulation of the motion of particles in a prescribed magnetic field geometry of a TCS was made in \cite{Eastwood1972}. This approach revealed first hints of the inhomogeneous distributions of plasma and electric current in the vicinity of TCS which may be associated with its fine structure. Later, such numerical study
approach evolved towards the application of more complex models and algorithms such as PIC \citep[]{Mingalev2007} and
Vlasov-Maxwell simulation \citep[]{Schmitz2006}. However numerical results not always clearly disclose the physical
nature of the simulated TCS features.

The analytical methods of kinetic description of plasma in TCS (developed in parallel with the numeric simulations) are related with solving of the nonlinear Vlasov-Maxwell equations \citep[]{Mahajan1989}. In general case, this is a complicated task. At the same time, it is known that the solution may be expressed via the invariants of motion. This fact forms a basis for various analytic approaches to the problem. For example, it was shown in \cite{Alexeev1970} that the magnetic momentum of a charged particle is conserved in the case of crossings of an infinitely thin current sheet. This result enabled to calculate the distributions of electric current and plasma in a simple prescribed magnetic field topology of a TCS \citep[]{Alexeev1990}.

The first self-consistent kinetic model of TCS was proposed by \cite{Kropotkin1996}. This model used the quasi-adiabatic invariant of proton oscillation across the TCS ($I_z=\oint V_z dz\approx const$) without taking into account of the electron component. Later on, the development of this approach enabled the inclusion of not only the electron component but also strong thermal anisotropy of plasma \citep[]{Zelenyi2004, Zelenyi2011, Artemyev2013}. It has to be mentioned that this self-consistent study of the TCS (based on quasi-adiabatic approximation) requires an extensive numerical calculations in order to obtain measurable macroscopic physical parameters of the TCS.

In the present paper we consider an easier analytical approach based on the methods applied in \cite{Alexeev1990}. It gives the consistent results which enable a comprehensive characterization of plasma and electric current distributions in the TCS without engagement of heavy numerical calculations. The considered simple one-dimensional case of the magnetic field geometry and the corresponding plasma particles motion may be attributed as a zero-order linear approximation of such natural phenomena as the current sheets of magnetotail, heliospheric and planetary magnetodiscs. In spite of the simplicity, the proposed approach enables characterization of the scaling features of the TCSs which, as it will be shown below, are consistent with the results obtained by other researches within more complicated modelling approaches. The presented study is based just on the analysis of plasma particles trajectories which are a fundamental issue of plasma physics.

The paper is organized as follows. In Section\,\ref{sec:Analytic solution} we describe our analytic solution based on the similar approach as in \cite{Alexeev1990}. In Section\,\ref{sec:Numerical calculation and analytical solution} we present the comparison of the obtained analytic solution and numeric calculations for the density of plasma and electric current. Section\,\ref{sec:scaling} discusses the issue of the typical scaling of the TCS obtained with our analytic solutions and shows its consistency with the known results of other authors. The paper findings are summarized and discussed in Section\,\ref{sec:Conclusion}.

\section{Analytic solution}\label{sec:Analytic solution}

The description of plasma motion in a strong magnetic field, when
the scale of the magnetic field variation is much larger than that
of the quasi-periodic motion of particles, is based on the
assumption of conservation of the adiabatic invariants. By this, the
thickness $d$ of a CS may be considered as a scale of
variation of the magnetic field, whereas the particle Larmor
rotation radii ($r_L^{p,e}$ for protons and electrons, respectively)
correspond to scale of quasi-periodic motion. For the case of
$\{r_L^{p,e}<d\}$ the motion of particles can be described by means
of the adiabatic theory of a guiding center using the decomposition
on a small parameter $r_L^{p,e}/d<1$. However, in a TCS an opposite condition $r_L^{p}\geq d$ usually holds true, and the adiabatic invariants are not conserved \citep[]{Cary1986, Sonnerup1971}, and for the description of particle motions in the TCS it is necessary to use a quasi-adiabatic theory \citep[]{Buechner1989, Zelenyi2011}.

The approach undertaken in this work is different from the attempts of a self-consistent description of a TCS in \cite{Kropotkin1996, Sitnov2000, Zelenyi2004}. We study the interaction of moving protons with the already existing super-thin current layer (STCL) with $\varepsilon=d/r_L<<1$, where $d$ stays for the thickness of the STCL and $r_L \equiv r_L^{p}$ is gyroradius of proton. This super-thin current layer is responsible for the existence of the prescribed background magnetic field. The goal is to study the distributions of the protons density and electric current in the vicinity of the STCL followed from the particle motion analysis. These distributions are related to the fine structure of a TCS. The physical nature of the postulated STCL could be related with the drifting electrons or with the cold  quasi-adiabatic protons.

The major point of the self-consistent model of TCS developed in \cite{Sitnov2000, Zelenyi2004} is that the electric current of a TCS and its maximal spatial scaling $\sim r_L$ (in case of weak anisotropy) are supposed to be controlled mainly by the same sort of particles (e.g. protons). Whereas in our case the presence of a STCL with $\varepsilon << 1$ created by another sort of particles (e.g., electrons or cold protons) is of principal importance. In that sense our approach is to certain extend analogous to the "matreshka" model by \cite{Zelenyi2006}.

The small parameter $\varepsilon \ll 1$ related with STCL allows to develop the adiabatic analytic approach for the analysis of proton motion. It is used in the asymptotic decomposition of physical quantities in the series over the powers of $\varepsilon$ (i.e. $\propto \varepsilon^n$, where $n=0,1,2..$). The vector velocity of proton in a zero order approximation ($n=0$) does not change the direction when a proton crosses the STCL. Therefore in a zero order approximation the STCL can be considered as an infinitely thin one.

The electric field of the charged particles can greatly complicate the analysis of their motion even in the zero
order of approximation. In that respect we consider a typical for the magnetospheric plasma case, when the Debye radius is several times less than the Larmor radius of proton ($r_L^{p} \equiv r_L = \frac{m_p \mid V \mid }{e \mid B \mid} \sin{\theta_0}$, where $\theta_0$ is the particle pitch-angle; $m_p$ and $e$ are mass and charge of proton, respectively), so that the effect of the particles own electric field is not important and the assumption of a quasi-neutrality of plasma holds true. In order to neglect the effects of the particle thermal motion, we consider the case when the energy of protons, entering the STCL, is much greater than the thermal energy, assuming for proton a simple spiral motion and taking electrons as an incompressible and massless liquid.


In our analysis we use an analog of the standard GSM coordinate
system with $Z$ - axis perpendicular to the STCL and $X$ - axis
directed along the tangential component of magnetic field (see in
Fig.\ref{fig1}a). For simplicity we assume the absence of a
large-scale electric field $E_y$ in the STCL which formally
corresponds to the transform into the de Hoffman-Teller frame. We also assume
that the variation of the normal component of magnetic field across
the infinitely thin current layer is small $\Delta B_z/\Delta Z \sim
0$. Hence from zero divergence of magnetic field the change of
tangential component along $X$ has to be also small, i.e. $\Delta
B_x/\Delta X \sim 0$. Such a configuration is widely used for the
analysis of most of the astrophysical phenomena, e.g. magnetotail
and magnetodisk.

Within the frame of the made geometry and coordinate system assumptions
the components of magnetic field are as follows: $B_{x}(z) =  B_{x}
sign(z)$; $B_{y}(z) = 0$; $B_{z}(z) = B_{z} = const$, so that
$\tan(\alpha)=B_{z}/B_{x}$ - characterizes the inclination of
magnetic field relative STCL. To distinguish between the physical quantities
in the $z>0$ and $z<0$ half-spaces we use further on for them the
indexes 1 and 2, respectively. We consider a monochromatic proton
beam which consists of particles having the same pitch-angle
$\theta$ and the same energy $\cal{E}$. In this case the whole beam
can be described based on the analysis of just one proton
trajectory. In other words, if the solution for one proton is known,
the behavior of the beam can be obtained by averaging of this
solution over $\theta$ and $\cal{E}$. By this, the trajectory of a
proton in the considered magnetic field geometry has a form of a
spiral, with the parameters which can be reconstructed from the
particle pitch-angle and phase of rotation of the proton at the moment before its
first crossing of the STCL (i.e., the boundary between regions 1 and
2).

After the first crossing of the STCL, the proton starts an oscillatory
motion across the current sheet (i.e., $XY$-plane), entering
consequently in the regions 1 and 2 separated by the STCL. During
this motion the proton makes a finite number, $n$, of crossing of the
STCL and during this process the particle pitch-angle gradually
changes (see Fig.\ref{fig1}b). Each pair of parameters
$(\theta_{1,2}^{m},\phi_{1,2}^{m})$ of the oscillating particle for
a particular $m-th$ crossing ($1 \leq m \leq n$) of the STCL is
uniquely defined by the values of these parameters at the previous
crossings (up to the very first entry into the STCL ($m=1$) due to the
recurrent character of the dependence). The lower index of the phase
and pitch-angle (i.e., 1 or 2) denotes the region from which the
particle approaches the STCL boundary. Note that for each crossing of
the STCL we distinguish between the phases and pitch-angles for the
particle entry in the current sheet and escape, $\phi_{1,2}^{m}$ and
$\theta_{1,2}^{m}$, which are marked by an appropriate combinations
of the indexes (see Fig.\ref{fig1}c).

From the conservation of the velocity vector at the STCL boundary the
following relations between $\theta_{1}^{m},\phi_{1}^{m}$ and
$\theta_{2}^{m},\phi_{2}^{m}$ hold true:

\begin{equation}
\cos(\theta_2^{m})=-\cos(\theta_1^{m})\cos(2\alpha)
-\sin(\theta_1^{m})\sin(2\alpha)\sin(\phi_1^{m}),
\label{eq1}
\end{equation}

\begin{equation}
\sin(\theta_2^{m})\cos(\phi_2^{m})=\sin(\theta_1^{m})\cos(\phi_1^{m}),
\label{eq2}
\end{equation}

\begin{equation}
\sin(\theta_2^{m})\sin(\phi_2^{m})=\cos(\theta_1^{m})\sin(2\alpha) -
\sin(\theta_1^{m})\cos(2\alpha)\sin(\phi_1^{m}).
\label{eq3}
\end{equation}

At the intervals between consequent crossings of the STCL the
pitch-angle of the particle remains unchanged. By this, the
following relation for the particle phase values for the consequent
crossings of the STCL may be obtained from the analysis of the
particle motion along its spiral trajectory segment  above (or
below) the STCL between the crossings

\begin{equation}
\phi_{1,2}^{m+1}+\frac{\tan(\theta_{1,2}^{m})}{\tan(\alpha)}\cos(\phi_{1,2}^{m+1}) = \phi_{1,2}^{m}+\frac{\tan(\theta_{1,2}^{m})}{\tan(\alpha)}\cos(\phi_{1,2}^{m}).
\label{eq4}
\end{equation}

In our analytical treatment we considered a special case of $\alpha\ll 1$. It corresponds to the field topology realized in magnetodisks, as well as in the magnetotail and some regions of the magnetopause. The smallness of $\alpha$ enables us to obtain the following approximate relations for $\theta_{1,2}^{m,m+1}$ and $\phi_{1,2}^{m,m+1}$ from equations (\ref{eq1})-(\ref{eq4})

\begin{equation}
 \theta_2^m=\pi-\theta_1^m+2\alpha\sin(\phi_1^m),
 \label{eq5}
\end{equation}

\begin{equation}
 \phi_2^m=2\pi-\phi_1^m+2\alpha\frac{\cos(\phi_1^m)}{\tan(\theta_1^m)},
 \label{eq6}
\end{equation}

\begin{equation}
 \phi_1^{m+1}=2\pi-\phi_1^{m}-2\frac{\alpha(\pi-\phi_1^m)}{\tan(\theta_1^m)\sin(\phi_1^m)}.
 \label{eq7}
\end{equation}

Using equations.(\ref{eq5}), (\ref{eq6}), (\ref{eq7}) the change of the
particle escape phase and pitch-angle at the STCL boundary
corresponding to one oscillation period (i.e. between $m$ and
$m+2$) may be calculated as

\begin{equation}
\Delta\phi_{1}^{m+2}=4\frac{\alpha}{\tan\theta_{1}^{m}}(\frac{(\pi-\phi_{1}^{m})}{\sin(\phi_{1}^{m})}+\cos(\phi_{1}^{m})),
\label{eq8}
\end{equation}

\begin{equation}
 \Delta\theta_{1}^{m+2}=4\alpha \sin(\phi_{1}^{m}).
\label{eq9}
\end{equation}

The same expressions can be written for the change of the particle
entry phase and pitch-angle $\Delta\phi_{2}^{m+2}$ and
$\Delta\theta_{2}^{m+2}$.

The ratio of the STCL escape (entry) phase to pitch-angle variations
on one oscillation period, i.e.
$\frac{\Delta\phi_{1,2}^{m+2}}{\Delta\theta_{1,2}^{m+2}}$ does not
depend on $\alpha$ (see equation (\ref{eq8}) and (\ref{eq9})). This fact,
in view of the smallness of $\alpha$ allows to transform this ratio
to the differential form:

\begin{equation}
\frac{d\phi}{d\theta}=\frac{1}{\tan(\theta)}\frac{(\frac{\pi-\phi}{\sin(\phi)}+\cos(\phi))}{\sin(\phi)}.
\label{eq10}
\end{equation}

Note that such transformation is possible only in the case of small jumps of $\Delta\phi_{1,2}^{m+2}$ and $\Delta\theta_{1,2}^{m+2}$ which implies an additional condition on $\alpha$ assumed initially just to be small. This condition is  $\alpha < \theta_0$, where $\theta_0$ is initial pitch-angle.

The solution of the equation (\ref{eq10}) is

\begin{equation}
\sin(\theta)= \sin(\theta_0) \frac{\sqrt{2(\pi-\phi_0)+\sin(2\phi_0)}}{\sqrt{2(\pi-\phi)+\sin(2\phi)}},
\label{eq11}
\end{equation}

where $\phi_0 \equiv \phi_{1}^{2}$ and $\theta_0 \equiv
\theta_{1}^{2}$ are the initial escape phase and pitch-angle of the
particle first oscillation cycle across the STCL. The same equation
as (\ref{eq11}) defines also the relation between the particle entry
phase and pitch-angle given $\phi_0$ and $\theta_0$ stay for their
initial values.

The equation (\ref{eq11}) can be also found from the simple analysis of
the trajectory shape and the related areas (full circle or truncated
circle) it covers (see Fig.\ref{fig2}\,a ). In the case when areas of the
full circle and the truncated circle are equal, the relation between
their radii $R_0$ and $R_1$ can be obtained from simple geometry
consideration: $R_1=R_0\sqrt{2 \pi}/\sqrt{2(\pi-\phi)+\sin(2\phi)}$.
On the other hand, taking into account the dependence of the
particle Larmor radius on the magnetic field and pitch-angle,
$R_1=m|V|\sin(\theta)/e|B|$, we obtain the analogous dependence of
$\sin(\theta)$ on the phase as in equation (\ref{eq11}). This result may be
interpreted as a conservation of the magnetic field flux through
particle Larmor circle area and then, the area inside a segment of
the particle trajectory between its consequent crossings of the STCL
\citep[]{Alexeev1970}.


In the analysis below we will use the following notations: $\phi$ -
for the phase of rotation of a particle in a homogeneous magnetic
field, i.e. on a segment of Larmor spiral between crossings of the
STCL; $\tilde \phi$ - for the escape phase of a particle, crossing
the STCL.

The density of particles with momentum in the range
$d\Gamma=p^2\sin(\theta)dp d\phi d\theta$ at infinity, is defined as

\begin{equation}
f(p,\theta,\phi)d\Gamma=f_0\delta(p-p_0)[\delta(\theta-\theta_0)+\delta(\theta-\pi+\theta_0)]p^2\sin(\theta)dpd\phi d\theta,
\label{eq12}
\end{equation}

where $f_0=n_0/(4\pi p_0^2 \sin(\theta_0))$, $n_0$ and
$p_0=\sqrt{2m{\cal E}}$ are density and momentum of particles,
respectively. The phase of rotation $\phi$ and escape phase $\tilde
\phi$ as the functions of coordinates can be calculated as follows. Expressing $V_z$ on the segment of trajectory ($\theta=\theta(\tilde\phi)$) in terms of $\phi$ and $\theta$ as $V_z=|V|(\cos(\theta)\sin(\alpha)-\sin(\theta)\sin(\phi)\cos(\alpha))$, taking into account the homogeneity of the magnetic field and the definition $V_z=dz/dt = \omega_L dz/d \phi$, where $\omega_L$ is Larmor frequency of proton, one can obtain by integration over $\phi$ from $\tilde\phi$ to $\phi$

\begin{equation}
z=\frac{r_L}{\sin(\theta_0)}[\sin(\theta(\tilde \phi))\cos(\alpha)(\cos(\phi)-\cos(\tilde\phi)) + \cos(\theta(\tilde \phi))\sin (\alpha) (\phi-\tilde \phi)].
\label{eq13}
\end{equation}

Introducing the projection of Larmor radius $r_L$ on the $z-$axis $z_0 =r_L\cos(\alpha)$, and taking into account the smallness of $\alpha$ the expression (\ref{eq13}) can be written as

\begin{equation}
z=z_0\frac{\sin(\theta(\tilde \phi))}{\sin(\theta_0)}[\cos(\tilde \phi)-\cos(\phi)].
\label{eq14}
\end{equation}

Substituting the dependence $\theta(\tilde \phi)$ defined by
equation (\ref{eq11}) into equation (\ref{eq14}) we obtain the relation for
$\phi$ to $\tilde \phi$:

\begin{equation}
\cos(\phi)=\cos(\tilde \phi)-\frac{z}{z_0}\sqrt{\frac{2(\pi-\tilde\phi)+\sin(2\tilde\phi)}{2(\pi-\tilde\phi_0)+\sin(2\tilde\phi_0)}}.
\label{eq15}
\end{equation}

Thus the distribution function (equation (\ref{eq12})) can be rewritten as

\begin{equation}
{\tilde f}(p,\theta,{\tilde \phi})d\Gamma=f_0\delta(p-p_0)[\delta(\theta-\theta(\tilde \phi))+\delta(\theta-\pi+\theta(\tilde \phi))]p^2\sin(\theta(\tilde \phi))\frac{\partial\phi}{\partial\tilde \phi}d\tilde\phi dp d\theta,
\label{eq16}
\end{equation}

where ${d\phi}/{d\tilde \phi}$ follows from (\ref{eq15}).

For integration of equation (\ref{eq16}) it is necessary to investigate
the boundaries of the phase space. In the considered model, two
major types of particle trajectory can be distinguished (see
Fig.\ref{fig1}). The first trajectory type corresponds to a simple
spiral rotation in a homogenous magnetic field, before the first and
after the last crossing of the STCL when particle approaches or
leaves the STCL. Another type of trajectory corresponds the
oscillatory motion of the particle across the STCL accompanied by the
particle gradual displacement due to the effect of $B_z$.

Therefore, each particle on its oscillatory part of the trajectory
appears to certain extend associated with the TCS and contributes
the specific near-by environment which may be considered as an
extension of  the TCS in the range $z = \pm z_0$. The range of
rotational phases of a particle approaching the STCL depends on its
z-coordinate. For $z \geq 2 z_0$ (i.e. the case of a pure spiral trajectory) the
rotational phase changes from 0 to $2 \pi$. For $z < 2 z_0$ the
phase of particles changes from $\phi_0(z)$ to $2\pi-\phi_0(z)$,
where $\phi_0(z)=\arccos(\cos (\tilde{\phi}_0) - z/z_0)$ and
$\tilde{\phi}_0$ is the phase of first crossing. By this,
$\tilde{\phi}$ changes during the particle oscillatory motion across
the STCL from ${\tilde \phi}_0$ till some maximum value
$\tilde\phi_m$ and then back to ${\tilde \phi}_0$ (Fig.\ref{fig2}\,b).
When the $\tilde{\phi}$ reaches its maximum value $\tilde\phi_m$,
the particle pitch-angle becomes $\pi/2$. In view of that one can
obtain from equation (\ref{eq11}) an equation which defines the value of
$\tilde\phi_m$:

\begin{equation}
2(\pi-\tilde\phi_m)+\sin(2\tilde\phi_m)=[2(\pi-\tilde\phi_0)+\sin(2\tilde\phi_0)]\sin^2(\theta_0).
\label{eq17}
\end{equation}

Average values of TCS (e.g., density and
electric current) in the close vicinity of STCL can be calculated as

\begin{equation}
<\eta(\theta_0,z)>= 2\int\limits_{\phi_0(z)}^{\pi-\phi_0(z)} \eta(\phi,\theta)d\phi+4\int\limits_{0}^{\tilde\phi_m}\eta(\tilde \phi,\theta(\tilde\phi),z)d\tilde\phi,
\label{eq18}
\end{equation}

where $\eta(\theta_0, z)$ corresponds to the particular physical
quantity which is averaged. The first integral in equation (\ref{eq18})
corresponds to the simple motion of the proton before first crossing.
The coefficient 2 before the first integral in equation (\ref{eq18}) is
due to the account of both types of protons -- the approaching and
leaving the STCL. The second integral is related to the protons'
periodical crossings of the STCL, whereas the factor 4 appears due to the
symmetry of the under-integral function on the intervals $\tilde\phi_0...\tilde\phi_m$ (or $\theta_0$ ... $\pi/2$) and $\tilde\phi_m...\tilde\phi_0$ (or $\pi/2$ ... $\pi-\theta_0$) and account of protons approaching to the TCS from both sides (e.g. from the regions 1 and 2). With formula (\ref{eq18}) one can calculate the physical
parameters of TCS for different values of pitch-angle $\theta_0$.

\section{Numeric simulation versus analytic solution}\label{sec:Numerical calculation and analytical solution}

To check the consistency of the obtained analytic solution we
compare it with the results of numeric simulations. Taking into
account that protons play the major role in defining the parameters
of the TCS and assuming their mutually independent motion,
we simulate numerically the dynamics of protons approaching
the STCL from the both sides.

In case, when the protons move independently from each other the trajectories may be calculated independently by numerical method. The standard 4-5 order Runge-Kutta numerical method has been used for solving the set of equations of motion. After that the calculated vector velocity (for each trajectory) was used to calculate the $\theta$ and $\phi$ for the corresponding coordinate $z$.

The goal of numerical simulations is to obtain the integral plasma parameters (according to equation (\ref{eq18})) defined with the reconstructed dependence $\theta=\theta(\phi)$ (where $\phi$ varies from $\phi_{min}$ to $\phi_{max}$) for fixed coordinate $z$. In course of test simulation runs we found that the amount of 30 protons is, in principle, enough to reconstruct $\theta=\theta(\phi)$ with a sufficient accuracy. To ensure better quality of the simulation and to minimize errors we finally take 60 protons approaching the STCL from the region 1 and another 60 protons coming to the STCL from an opposite side, i.e. the region 2. The module of initial velocity was assumed to be the same for all particles (i.e. the monochromatic case was considered).

To calculate the particle trajectories we used the normalized equation of motion

\begin{equation}
\frac{d^2{\bf r}}{dt^2}= [{\bf V}\times{\bf B}].
\end{equation}

In course of the simulation, we use the same coordinate system and
the geometry of magnetic field as those in the case of analytic
treatment and take for the calculations $B_z=0.01$ and
$B_x(z)=sign(z)$. The initial velocity of particles was taken ${\bf
V}=\{V_{\parallel}, V_{\perp1}, V_{\perp2} \}$ and $|V|=1$ with the components
depending on the initial values $\theta_0,\phi_0$ as follows

\begin{equation}
 V_{\parallel}= \cos(\theta_0),
\end{equation}

\begin{equation}
 V_{\perp1}= - \sin(\theta_0)\cos(\phi_0),
\end{equation}

\begin{equation}
 V_{\perp2}= \sin(\theta_0)\sin(\phi_0).
\end{equation}

By this, for the regions 1 ($z>0$) and 2 ($z<0$) the initial
pitch-angle values for all particles were taken as $\pi-\theta_0$
and $\theta_0$, respectively. The initial phases for different
particles approaching the STCL in the same region (i.e. in the region
1 or 2) were taken as $\phi_{0}=2\pi/n$, where $n$ is number of the
particle ($n=1,2,...,60$).

The initial position of the particles
guiding centers were $8r_L$ and $-8r_L$ for the particles in the
regions 1 and 2, respectively. The size of the simulation box in
z-direction was $z=\{-9r_L;9r_L\}$.

Based on all calculated 120 trajectories the distribution function equation (\ref{eq12})
was reconstructed in the parameter space $\{\theta,\phi\}$ for the
running values of $z-$coordinate. Finally the electric current $J_y$
and number density $N$ as the functions of $z$ are given by the
integrals:

\begin{equation}
 J_{y}(z)=J_0\int\limits_{\phi_{min}(z)}^{\phi_{max}(z)}\sin^2(\theta(\phi))\cos(\phi)d\phi,
\end{equation}
\begin{equation}
 N(z)=N_0\int\limits_{\phi_{min}(z)}^{\phi_{max}(z)}\sin(\theta(\phi))d\phi,
\end{equation}

where $\phi_{min}(z)$ and $\phi_{max}(z)$ stay for the boundaries of
the considered phase-space for the corresponding $z$. The dimensional coefficients $N_0,J_0$ have the following form $N_0={\tilde N}_0/4\pi\sin(\theta_0), J_0=eV_0N_0$ where $V_0$ and ${\tilde N}_0$ are module velocity and the number density of incoming flow, respectively.

To investigate the influence of the particle pitch-angle on the
electric current and particle number density profiles of TCS along $z$ we
perform calculations for three different values of
$\theta_0=\{0.2,0.4,0.6\}$. Due to the symmetry of the problem (in
the regions 1 and 2), the calculation results for only one
half-space (e.g., $z\geq 0$) are presented in Fig.
\ref{fig3}.


Both, the particle number density and the electric current reach the
maximum value not in the center of the current sheet ($z=0$), but
slightly aside of it (at $z=z_m$). Similar effect was discovered also
in \citep[]{Eastwood1972}. The obtained splitting (i.e. shifted
maximum) of the electric current and particle density profiles is
due to the oscillatory character of particle motion in the close
vicinity ($\pm z_m$) of the TCS. Note, that in the case of
Maxwellian velocity distribution of particles approaching the TCS
the effect of splitting becomes less pronounced than in the above
considered monochromatic case.

As one can see in Fig.\ref{fig3}, the
results of the numeric simulation are rather close to the analytic
solutions given by the equation (\ref{eq18}). However, there is some
difference in the amplitudes of maxima of the electric current and
particle density. That is because in obtaining our analytic
solution we considered the approximate equations (\ref{eq14}) and
(\ref{eq15}) which ignore the terms of the order of $\alpha \ll 1$.
Therefore, the discrepancy between the numerical and analytic
solutions is of the order of $\delta\sim \alpha/\tan(\theta_0) \sim
B_z/(B_x \tan(\theta_0))$ which yields $\delta\sim 0.05$ and
$\delta\sim 0.025$ for $\theta_0=0.2$ and $\theta_0=0.4$,
respectively. Besides of that, certain difference between the
analytic and numeric solutions is related to different account the first
phase ${\tilde \phi}_0$ of crossings of protons. The analytic solution was obtained for a particular value
${\tilde \phi}_0$ of the entry phase, the same for all particles, whereas in
the numeric simulation the particle entry phases are different for
different particles. In order to compare the numerical and
analytical results, an average entry phase for the first crossing of
STCL for the whole ensemble of model particles was used as $\phi_0$
in the analytical solution.

Note, that dimensionless plasma density $N_{\infty}=N(z>2r_L)/N_0$ outside the TCS, according
both the analytic solution and numeric simulation, shows the
following dependence on the pitch-angle: $4\pi\sin(\theta_0)$,
whereas the ratio of the TCS density maximum $N_{max}=N(z_{m})/N_0$ to
$N_{\infty}$ decreases for the increasing $\theta_0
=\{0.2,0.4,0.6\}$ as $N_{max}/N_{\infty}=\{1.55,1.25,1.18\}$ (where $z_{m}$ is coordinate of maximal value of $N/N_0$). Also the ratio of the plasma density $N_{z=0}=N(0)/N_0$ in the center of the TSC
to $N_{\infty}$ decreases: $N_{z=0}/N_{\infty}=\{1.36,0.93,0.76\}$
and can even reach the values less than one. Therefore, we see that
the particles with different pitch-angles contribute differently to
the density distribution in the TCS and its vicinity. Decrease of
$\theta_0$ results in the compression of the TCS, whereas the
increase of $\theta_0$ produces strong rarefication of the number
density in the center ($z=0$).

Besides of growing $z_m$, also the value of the electric current maximum increases for the increasing
pitch-angle: $J_{max}=J(z_m)/J_0=\{1.32,1.41,1.45\}$. Note that the sign of the
current in the center of the TCS ($z=0$) changes for the growing
$\theta_0$: $J_{z=0}=J(0)/J_0=\{0.6,-0.1,-1.2\}$. This is because the
particles with sufficiently big $\theta_0$ create (because of their
trajectory form) a negative electric current
which is opposite to the main current.

\section{The scaling of TCS}\label{sec:scaling}

Based on the above presented analytic approach, it is possible to
estimate the value of the characteristic scale $z_S=z_m/r_L$ of the
TCS fine structure (where $z_m$ is coordinate of the electric current and plasma density maximum)
and its dependence on the pitch-angle $\theta_0$. We pay
attention to the fact that the particles oscillating across the STCL
in its close vicinity move along the quasi-circular trajectories
(see Fig.\ref{fig1}), so that the particles with higher $y-$
component of velocity give the higher contribution to the electric
current. The same is true for the particle contribution to the
number density. By this, according to the equations (\ref{eq2}) and
(\ref{eq14}), $V_y$ is a function of $\tilde{\phi}$,
$\theta(\tilde{\phi})$ and rotational phase $\phi$. Therefore, the
location of electric current maximum approximately corresponds
$\phi=\pi$ and $\theta({\tilde\phi}_{m})= \pi/2$. In view of that, and
assuming for the definiteness sake ${\tilde\phi}_0=0$, we obtain from the
equation (\ref{eq14}) an expression for the estimation of $z_S$:

\begin{equation}
 z_S= \frac{1+\cos({\tilde \phi}_m)}{\sin(\theta_0)},
 \label{eq19}
\end{equation}

where  ${\tilde \phi}_m$ is the solution of equation (\ref{eq17}). The
Fig.\ref{fig4} illustrates the dependence of $z_S$ on the
pitch-angle $\theta_0$ as provided by the equation (\ref{eq19}). In
particular, for the increasing $\theta_0$ the coordinate of maximum
of the electric current and plasma density in the TCS shifts towards
larger $z$ up to the maximum value $2r_L$. That may be considered as
a broadening of the TCS. Note that our analysis is valid under the assumption of
$\alpha < \theta_0$, therefore approaching to $\theta_0 = 0$ corresponds the tangential discontinuity and as result the self-thickness became zero.


In process of crossings the pitch-angle is growing (see. Fig. \ref{fig2}b) and as result the maximal Larmor radius ($\theta=\pi/2$) can be expressed as $r_{L}^{max}=r_{L}/\sin(\theta_0)$.
In this case the expression for the characteristic scale equation (\ref{eq19}) yields $z_m/r_{L}^{max}={1+\cos({\tilde \phi}_m)}$, where ${\tilde \phi}_m$ is a function of $\theta_0$. Taking that the pitch-angle can be expressed as $\theta_0=\arctan(V_{\perp}/V_{\parallel})$ and considering the case of a strongly anisotropic plasma beam (i.e., $V_{\perp}\rightarrow 0$) we obtain for the TCS scaling the well known relation $z_m/r_{L}^{max}={(V_{\perp}/V_{\parallel})^{4/3}}$. This scaling was found first in \cite{Francfort1976} within frame of a similar to our approach. Considering the TCS scaling in terms of $r_{L}$ in the similar case of strong anisotropy one can obtain from equation (\ref{eq19}) that $z_m/r_{L}={(V_{\perp}/V_{\parallel})^{1/3}}$. The same result was reported in \cite{Sitnov2000} in the limiting case $B_n/B_0<V_t/V_D<<1$ (where $V_t,V_D$ are thermal and flow velocities) which is also supposed to hold true ($\alpha < \theta_0$) in our model for case of $\theta_0\ll 1$.

The presented analytic approach enables clear interpretation of the different TCS scaling types obtained in \cite{Francfort1976} and \cite{Sitnov2000}. In particular, the power index $4/3$ corresponds the scaling in terms of the TCS internal Larmor radius (i.e. $r_{L}^{max}$), whereas index $1/3$ appears in the scaling defined in terms of the incoming proton Larmor radius ($r_{L}$).

Our method enables a generalization for the case when the protons have certain distribution over velocities and pitch-angles ($\theta_0$), which is described by the distribution function $f(V,\theta_0)$. The average half thickness of the TCS $<z_m>$ in this case can be defined as follows

\begin{equation}
 <z_m>= \frac{\int\limits_{0}^{\infty}dV\int\limits_{0}^{\pi/2}d\theta_0\,\, z_m(V,\theta_0) f(V,\theta_0)V^2 \sin(\theta_0)}{\int\limits_{0}^{\infty}dV\int\limits_{0}^{\pi/2}d\theta_0\,\, f(V,\theta_0)V^2 \sin(\theta_0)},
 \label{eq20}
\end{equation}

where $z_m(V,\theta)=\frac{V}{\omega_L}(1+\cos({\tilde \phi}_m(\theta_0)))$. Note that the dependence of ${\tilde\phi}_m(\theta_0)$ is given by equation (\ref{eq17}), where $\tilde\phi_0=0$. For simplicity, we consider here a particular class of the distribution function of the following form $f(V,\theta_0)=f_v(V)f_{\theta}(\theta_0)$, assuming $f_{v}=e^{\frac{-(V_D-V)^2}{2V_t^2}}$. The last corresponds the shifted Maxwellian distribution with the flow velocity $V_D$ and thermal velocity $V_t$. Finally, the equation (\ref{eq20}) can be rewritten as,

\begin{equation}
 <z_m>=\sqrt{2}\,\rho_{t}\, F[\frac{V_D}{\sqrt{2}V_t}]\,\{1+\frac{\int\limits_{0}^{\pi/2} \cos({\tilde\phi}_m(\theta_0))f_{\theta}(\theta_0)\sin(\theta_0)d\theta_0}{\int\limits_{0}^{\pi/2} f_{\theta}(\theta_0)\sin(\theta_0)d\theta_0}\},
 \label{eq21}
\end{equation}

where $\rho_{t}=\frac{V_t}{\omega_L}$ is thermal Larmor radius and the function $F[\frac{V_D}{\sqrt{2}V_t}]=\frac{\int\limits_{0}^{\infty}\zeta^3e^{-(\frac{V_D}{\sqrt{2}V_t}-\zeta)^2 }d\zeta}{\int\limits_{0}^{\infty}\zeta^2e^{-(\frac{V_D}{\sqrt{2}V_t}-\zeta)^2}d\zeta}$ can be expressed via the tabulated error integral function ($erf(V_D/\sqrt{2}V_t) = \frac{2}{\sqrt{\pi}} \int\limits_{0}^{V_D/\sqrt{2}V_t} \exp{-\zeta^2} d\zeta$). The equation (\ref{eq21}) describes the dependence of thickness of TCS in terms $\rho_t$ as a function of $V_D/\sqrt{2}V_t$.

To investigate the dependence of the scaling of TCS on the incoming proton flow velocity $V_D$ we rewrite equation (\ref{eq21}) as,

\begin{equation}
 \frac{<z_m>}{\rho_{t}} =\sqrt{2}\,F[\frac{V_D}{\sqrt{2}V_t}]\,\, \cal{K},
 \label{eq22}
\end{equation}

where $\cal{K}$ stays for the finite expression in brackets in the equation (\ref{eq21}). The behavior of the function $F[\frac{V_D}{\sqrt{2}V_t}]$ is presented on fig.\ref{fig5}. In this figure one can see that for large values of $V_D/\sqrt{2}V_t$ ($\geq 1.5$) the function $F[\frac{V_D}{\sqrt{2}V_t}]$ becomes a linear one (proportional to $\frac{V_D}{\sqrt{2}V_t}$). In view of that the scaling expression equation (\ref{eq22}) can be used for the estimation (with accuracy up to the proportionality coefficient) of energy of the incoming proton flow resulting in particular half thickness of TCS, or vice versa, to calculate the thickness of the TCS based on a given energy of the flow, using the definitions $V_t=\sqrt{T/m_p}$ and $V_D\sim \sqrt{2{\cal E}/m_p}$ where $\cal E$ is the energy of the flow:

\begin{equation}
 \frac{<z_m>}{\rho_{t}} \propto \sqrt{\frac{2{\cal E}}{T}}\,\, \cal{K},
 \label{eq23}
\end{equation}

where $T$ is temperature of the incoming proton flow measured in energy units. Specifically, in the case of a proton beam with an energy ${\cal E}\sim 12$ keV and thermal dispersion corresponding $T \sim 1.5$ keV it follows from equation (\ref{eq23}) that $\frac{<z_m>}{\rho_{t}} \propto 4 \cal{K}$. If the distribution function $f_{\theta}(\theta)$ turns ${\cal K}$ to $\sim 1$, then the scale of half thickness of TCS in this particular case is $\sim 4$. Such typical scales are often measured for the TCSs in the terrestrial magnetotail  \citep[]{Runov2006}.



\section{Conclusion}\label{sec:Conclusion}

In this paper we present the results of analytic treatment of proton dynamics in the vicinity of TCS and compare them with the corresponding numerical modeling, without taking into account of the effect of electrons. Both, numerical and analytic results appear in good qualitative and quantitative agreement. That demonstrates the high efficiency and potential usefulness of the proposed analytical approach for the analysis of TCSs.

It was numerically shown in \cite{Harold1996} that the negative component of electric current (diamagnetic current) could play significant role in case $V_D<<V_t$. The role of dia- and paramagnetic currents was studied theoretically in \cite{Zelenyi2000}. In particular, in the case of $V_D>>V_t$, the diamagnetic effects are proportional to $V_\perp^2$ and became small due to smallness of $V_t$. Our results demonstrate similar picture, where the negative  (diamagnetic) current increases with the increasing pitch-angles (i.e. decreasing $V_D/V_t$) and vice versa. Also the number density in the TCS strongly depends on the value of the pitch-angle and the resulting density distribution in the TCS is connected with the value of particles pitch-angle, which may (depending on particular parameters) look as a density increase or rarefication.

Based on the analytic approach it was possible to interpret the obtained distribution of the electric current and particle density near the neutral line and to estimate its characteristic scale $z_S$ as a function of particle pitch-angle. The obtained analytic expression for $z_S$ has been shown to agree in the limiting cases with known estimates of the TCS scaling derived under more general assumptions \citep{Francfort1976, Sitnov2000}. Within the frame of the presented analytic approach it was possible to give an interpretation of the different TCS scaling types (i.e., the power indexes $4/3$ and $1/3$) which are connected with the scaling in units of differently  defined Larmor radii. Finally, the estimations of the average scaling of the half-thickness of TCS, with the inclusion of the thermal dispersion of the incoming proton beam gave for the particular (common) parameters of the beam the values $\frac{<z_m>}{\rho_{t}} \approx 4$, which are often observed in the Earth's magnetotail.

The proposed analytical approach may be efficient in further investigation of TCS fine structure in the more general case. In particular, as further development of the presented study may be construction of self-consistent solutions for the magnetic field and electric current by an appropriate adjustment of the proton flow and particle pitch-angle distribution function.

\section*{Acknowledgments}\label{sec:acknowl}
This work was supported by the Austrian Science Foundation (FWF)
(project S11606-N16). The authors acknowledge the EU FP7 project IMPEx
for providing collaborative environment for research and communication.

M. O. V. acknowledges the financial support of Division of Physical Sciences of the Russian Academy of Sciences (program VI.15 �Plasma processes in Space and in Laboratory�).

For more details on the computer code used in the paper the reader is invited to contact the first author (jury.sasunov@oeaw.ac.at).


\end{article}

\pagebreak
\renewcommand{\thefigure}{\arabic{figure}}

\begin{figure}
\begin{minipage}[h]{0.29\linewidth}
\center{\includegraphics[width=1\linewidth]{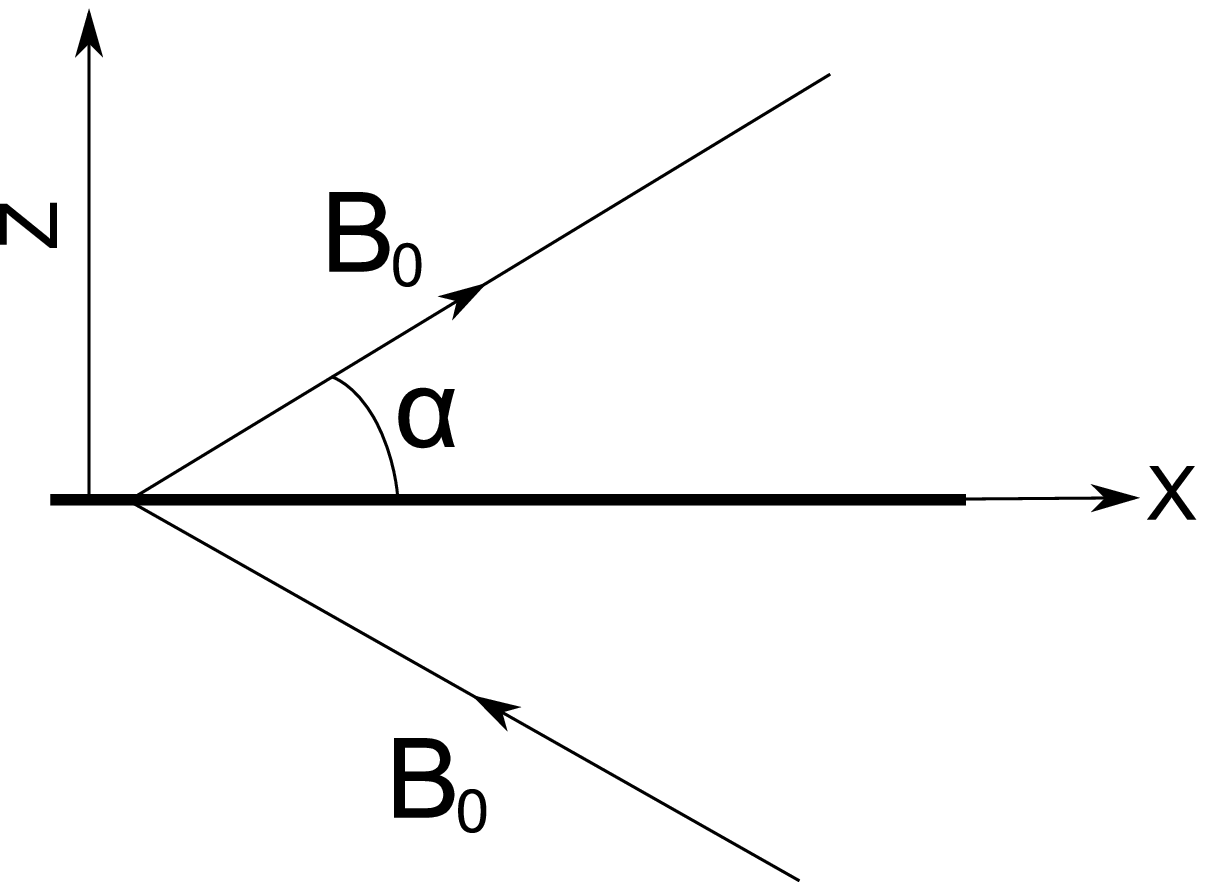}} a) \\
\end{minipage}
\begin{minipage}[h]{0.29\linewidth}
\center{\includegraphics[width=1.5\linewidth]{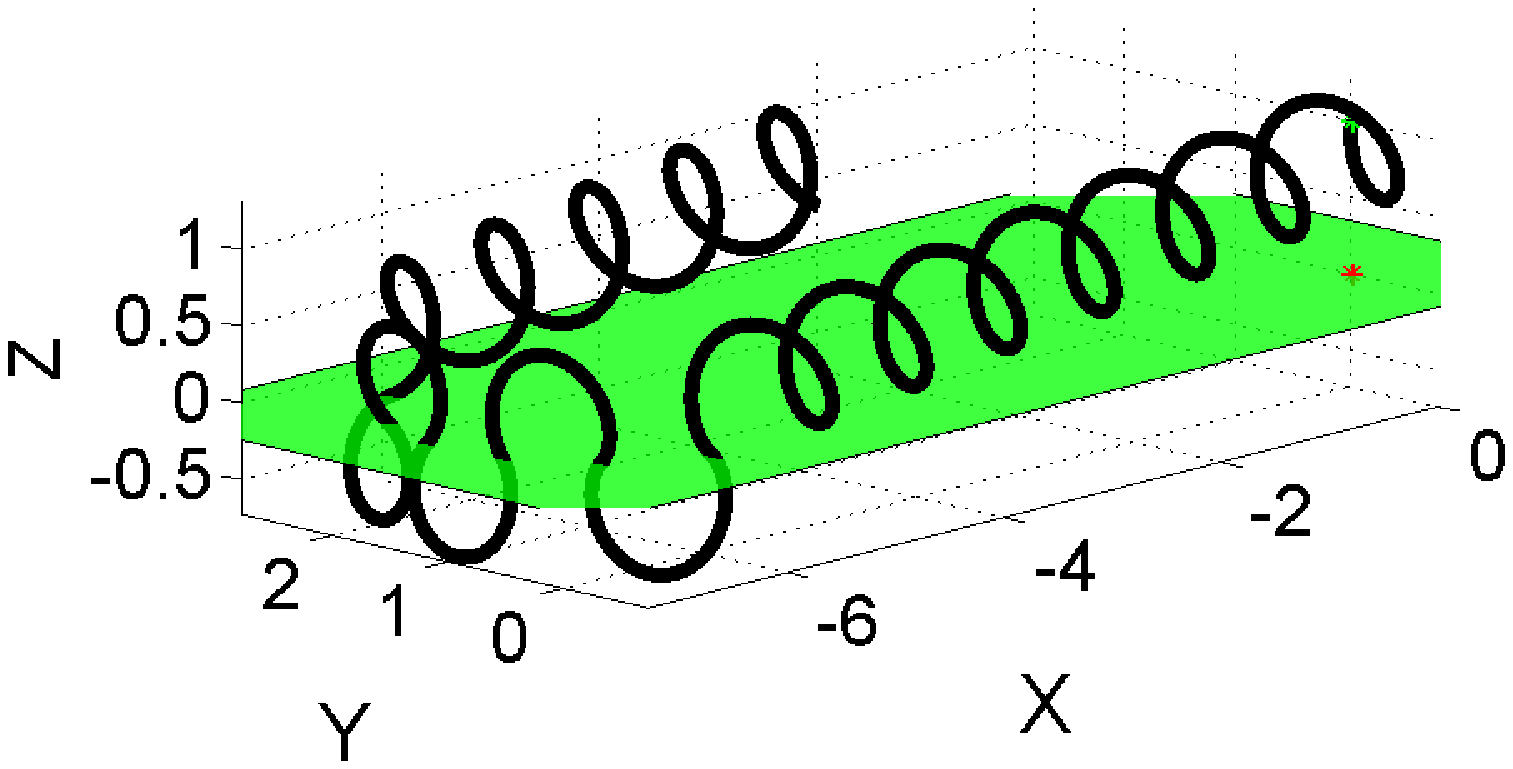}} b) \\
\end{minipage}
\vfill
\begin{minipage}[h]{0.35\linewidth}
\center{\includegraphics[width=1.5\linewidth]{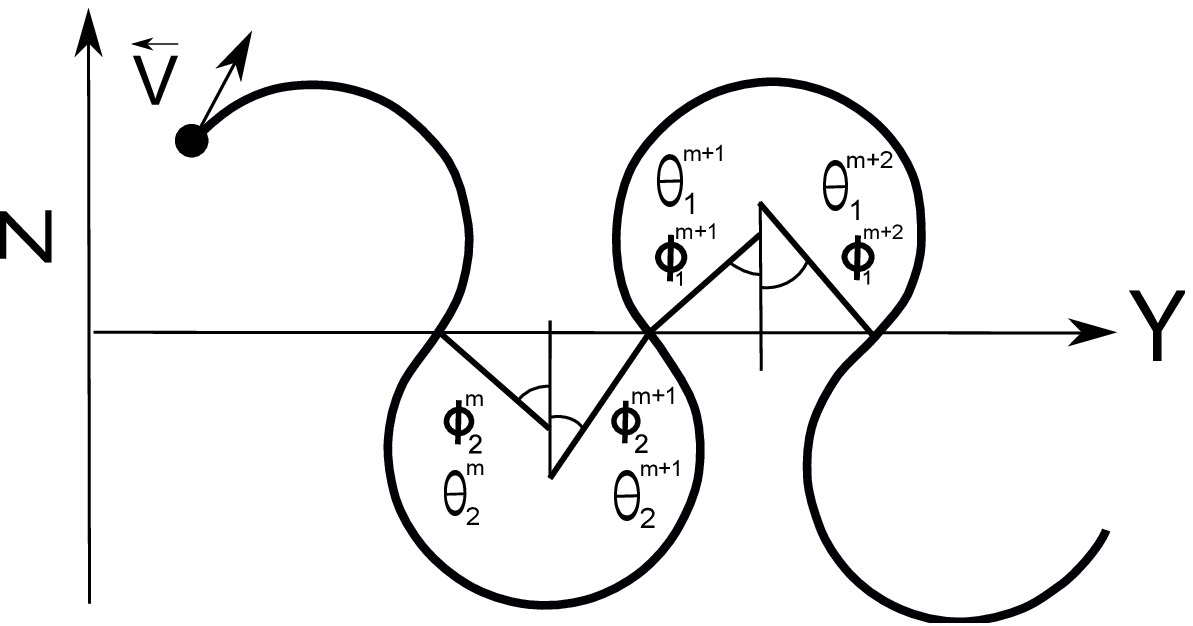}} c) \\
\end{minipage}
\caption{The configuration of the magnetic field (a); A schematic view of the proton trajectory oscillating around the TCS with the entry and escape phases denoted for each crossing of the TCS (b,c). The lower index of the phase and pitch-angle (i.e., 1 or 2) denotes the region from which the particle approaches the TCS, the upper index of phase denotes number of crossing. }
\label{fig1}
\end{figure}

\begin{figure}
\begin{minipage}[h]{0.5\linewidth}
\center{\includegraphics[width=1\linewidth]{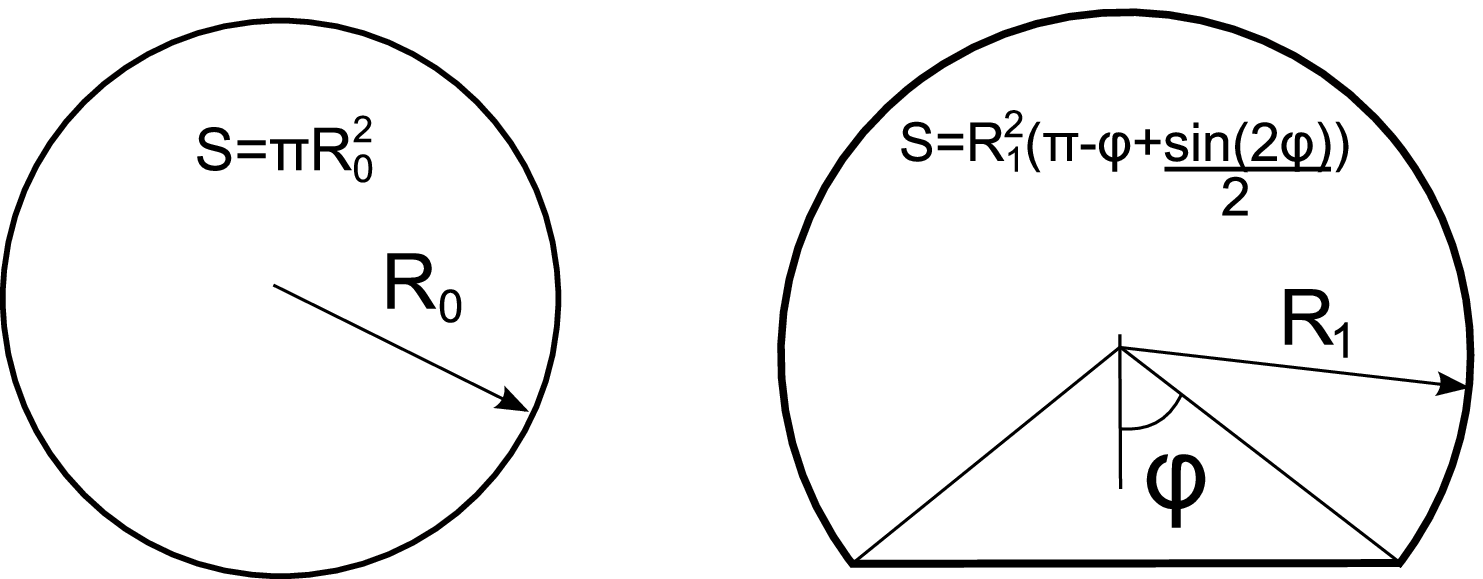}} a) \\
\end{minipage}
\begin{minipage}[h]{0.29\linewidth}
\center{\includegraphics[width=1\linewidth]{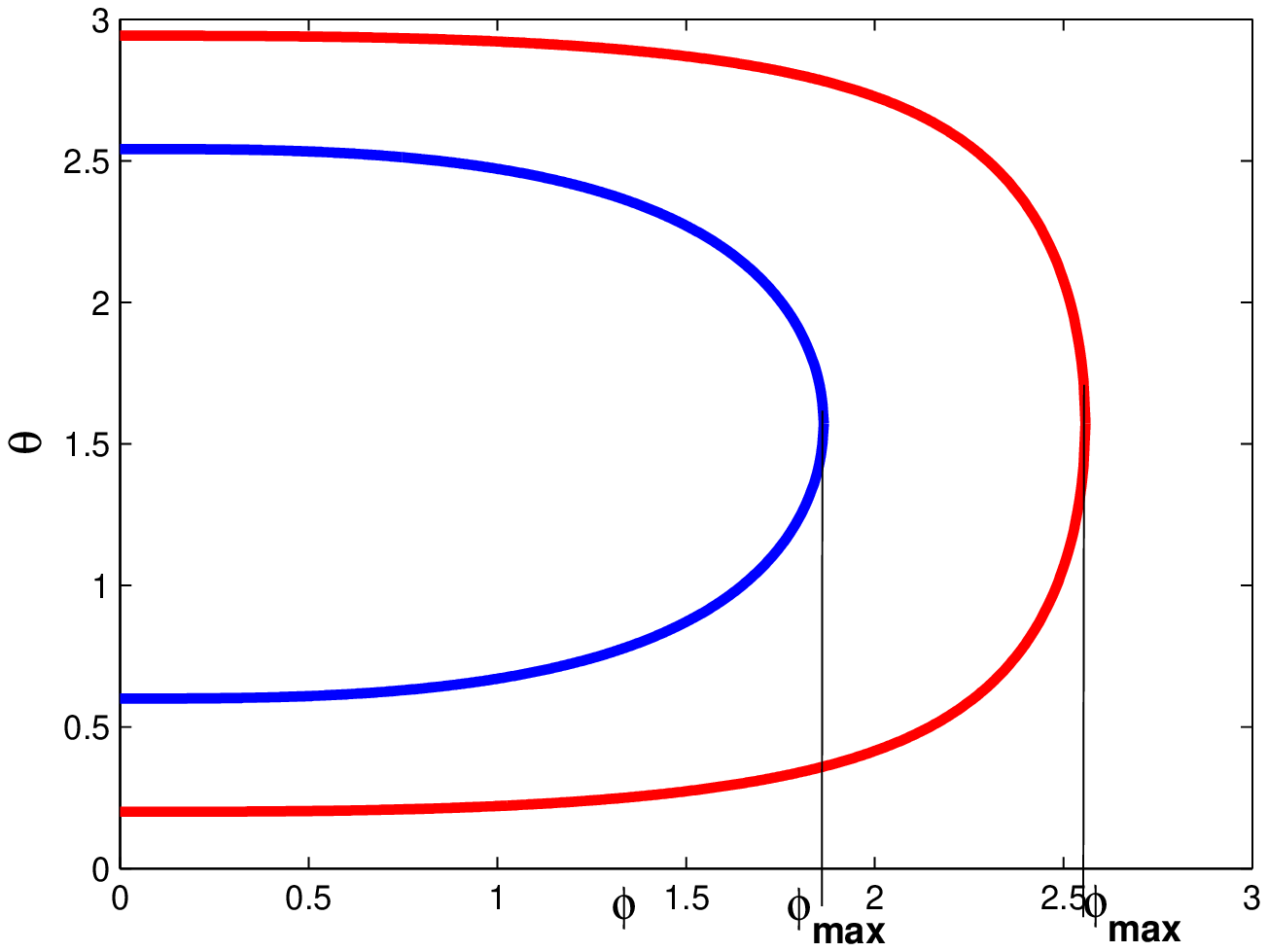}} b) \\
\end{minipage}
\caption{The areas of the full and truncated circles covered by the particle trajectory in the case of free spiral motion and oscillation across the TCS, respectively a). The behavior of pitch-angle $\theta(\phi)$ as a function of phase for different values of $\theta_0$ b). The red and blue lines correspond $\theta_0=0.2$ and $\theta_0=0.6$, respectively. The phase $\phi_{max}$ is a maximum phase for which particle may cross the TCS. }
\label{fig2}
\end{figure}

\begin{figure}[H]
\begin{minipage}[h]{0.47\linewidth}
\center{\includegraphics[width=1\linewidth]{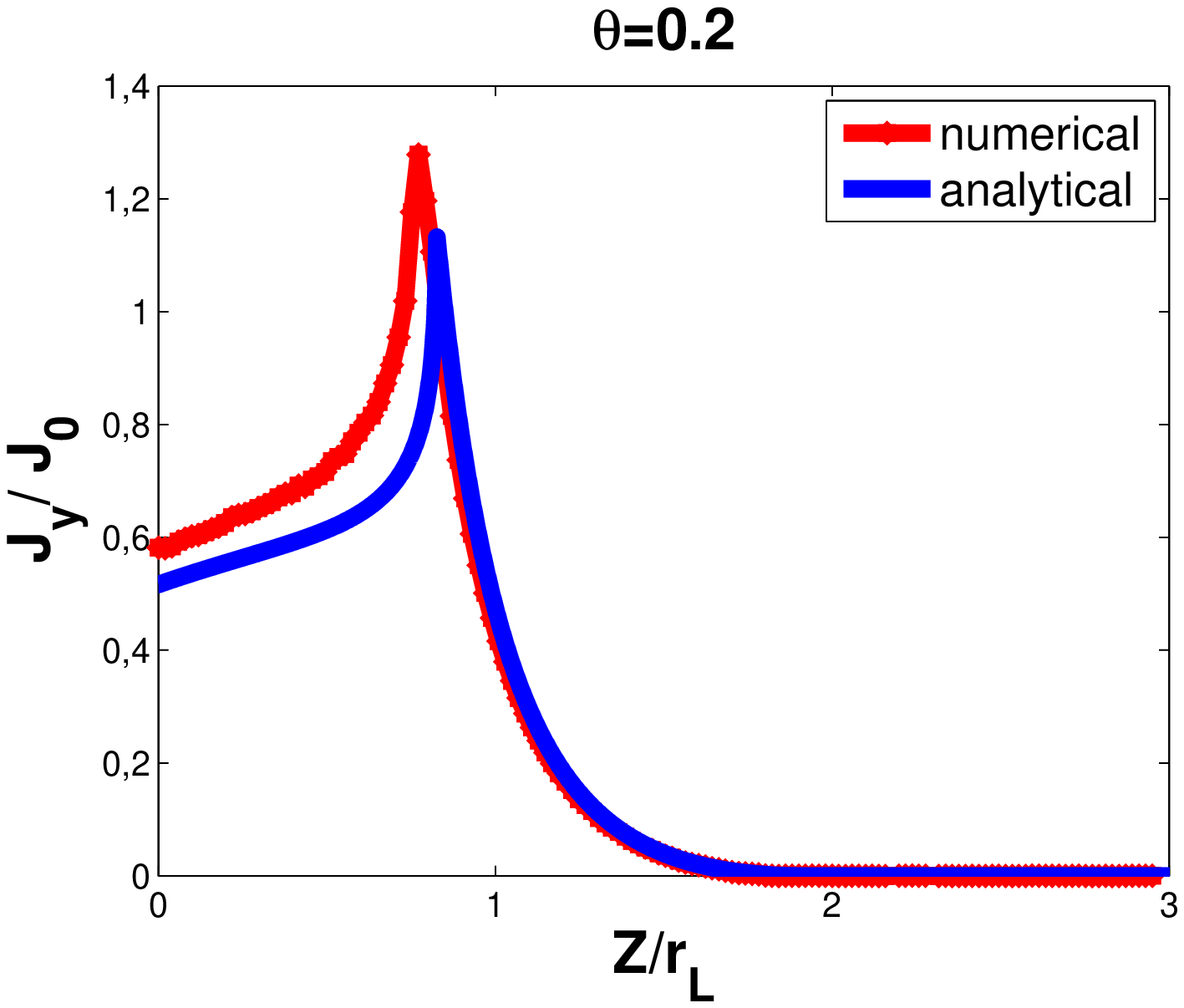}} a) \\
\end{minipage}
\hfill
\begin{minipage}[h]{0.47\linewidth}
\center{\includegraphics[width=1\linewidth]{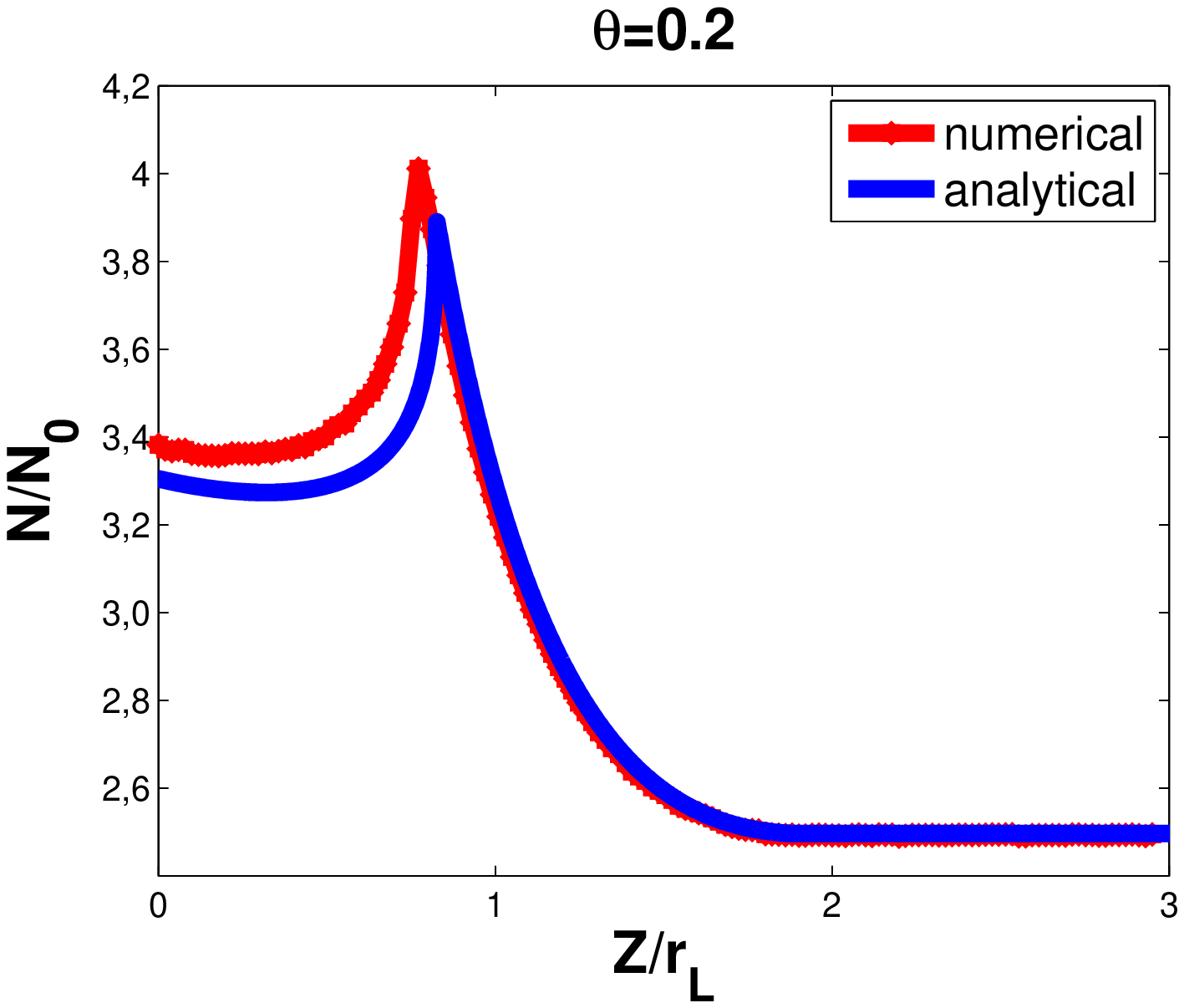}} \\b)
\end{minipage}
\vfill
\begin{minipage}[h]{0.47\linewidth}
\center{\includegraphics[width=1\linewidth]{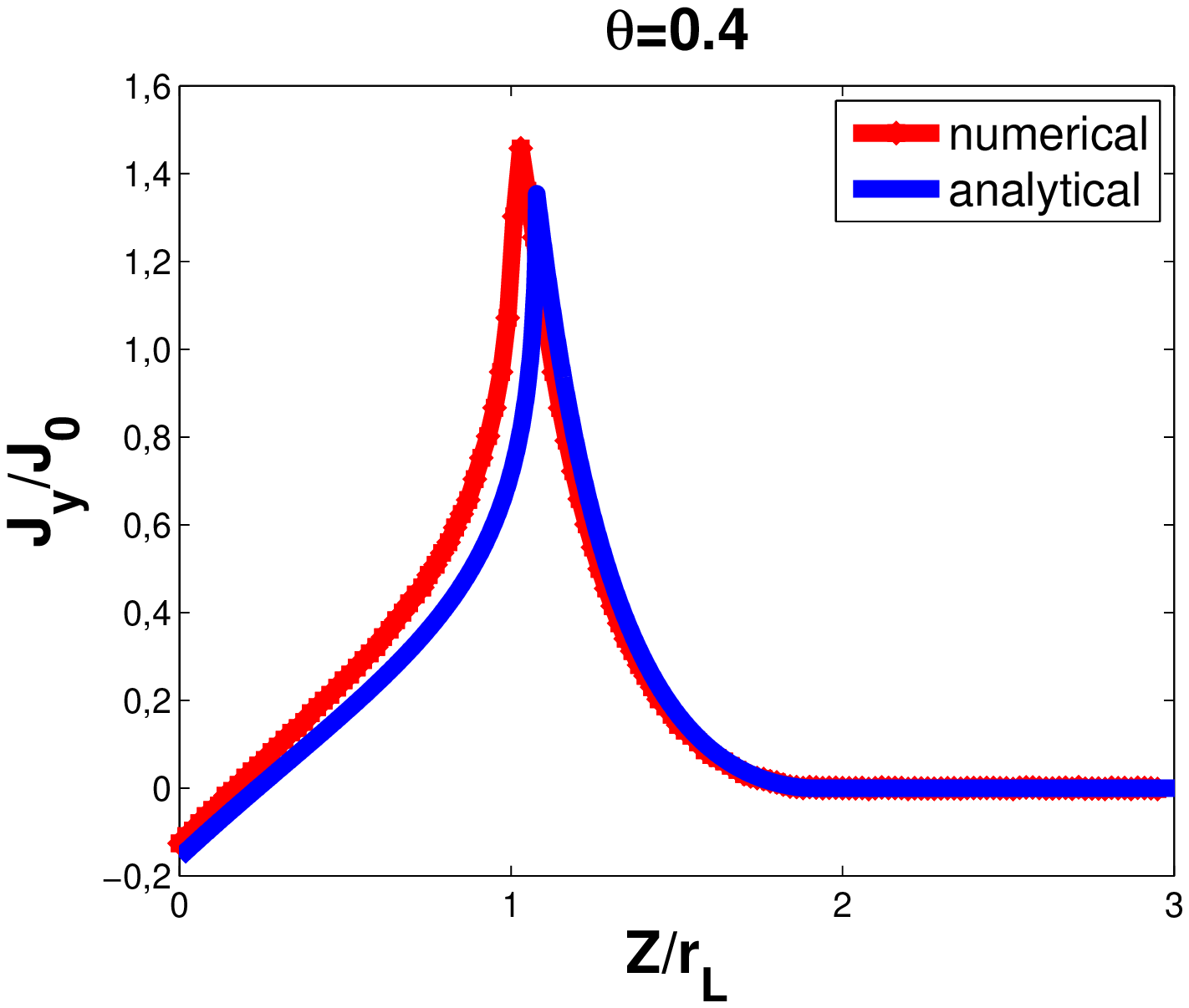}} c) \\
\end{minipage}
\hfill
\begin{minipage}[h]{0.47\linewidth}
\center{\includegraphics[width=1\linewidth]{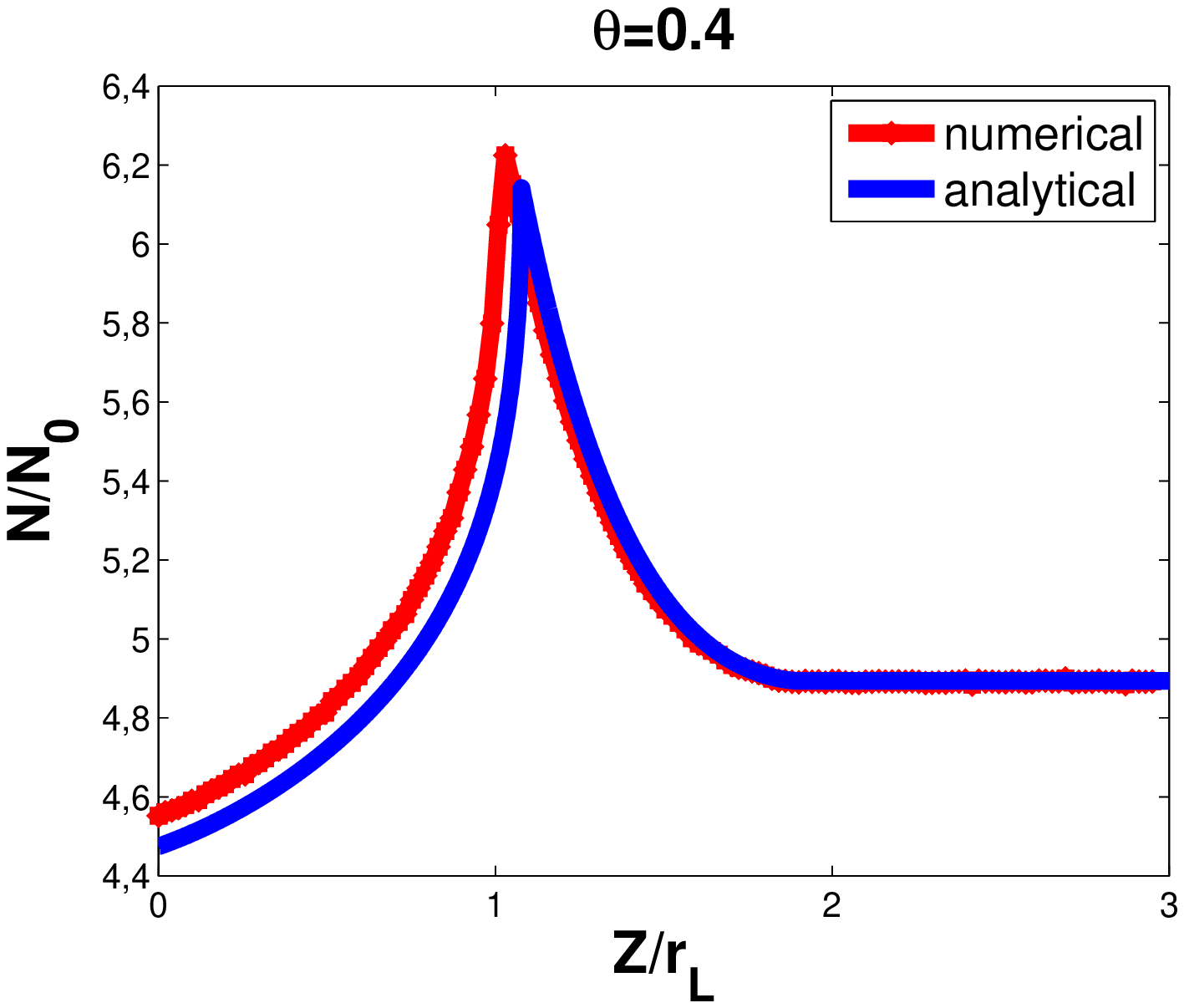}} d) \\
\end{minipage}
\vfill
\begin{minipage}[h]{0.47\linewidth}
\center{\includegraphics[width=1\linewidth]{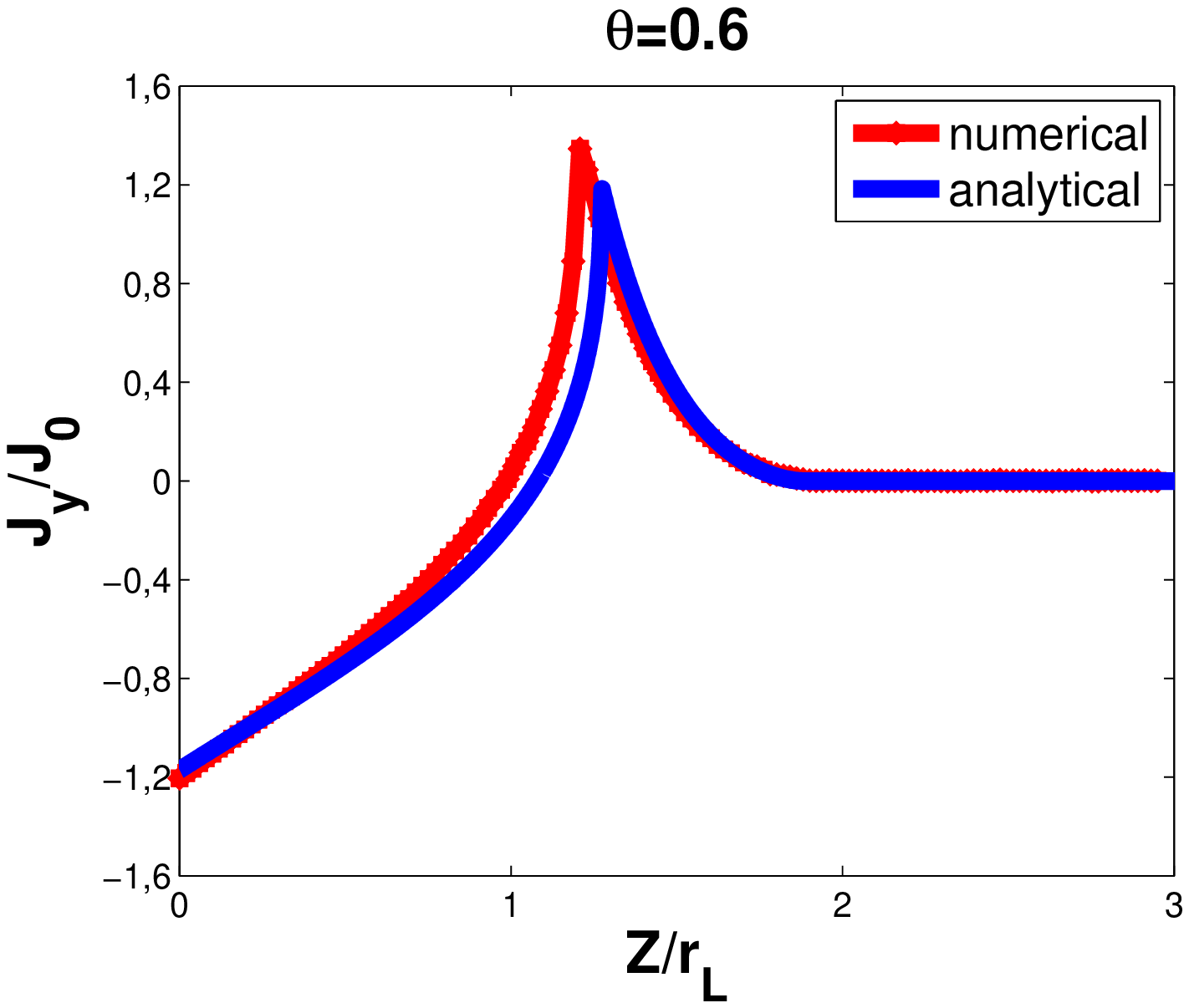}} e) \\
\end{minipage}
\hfill
\begin{minipage}[h]{0.47\linewidth}
\center{\includegraphics[width=1\linewidth]{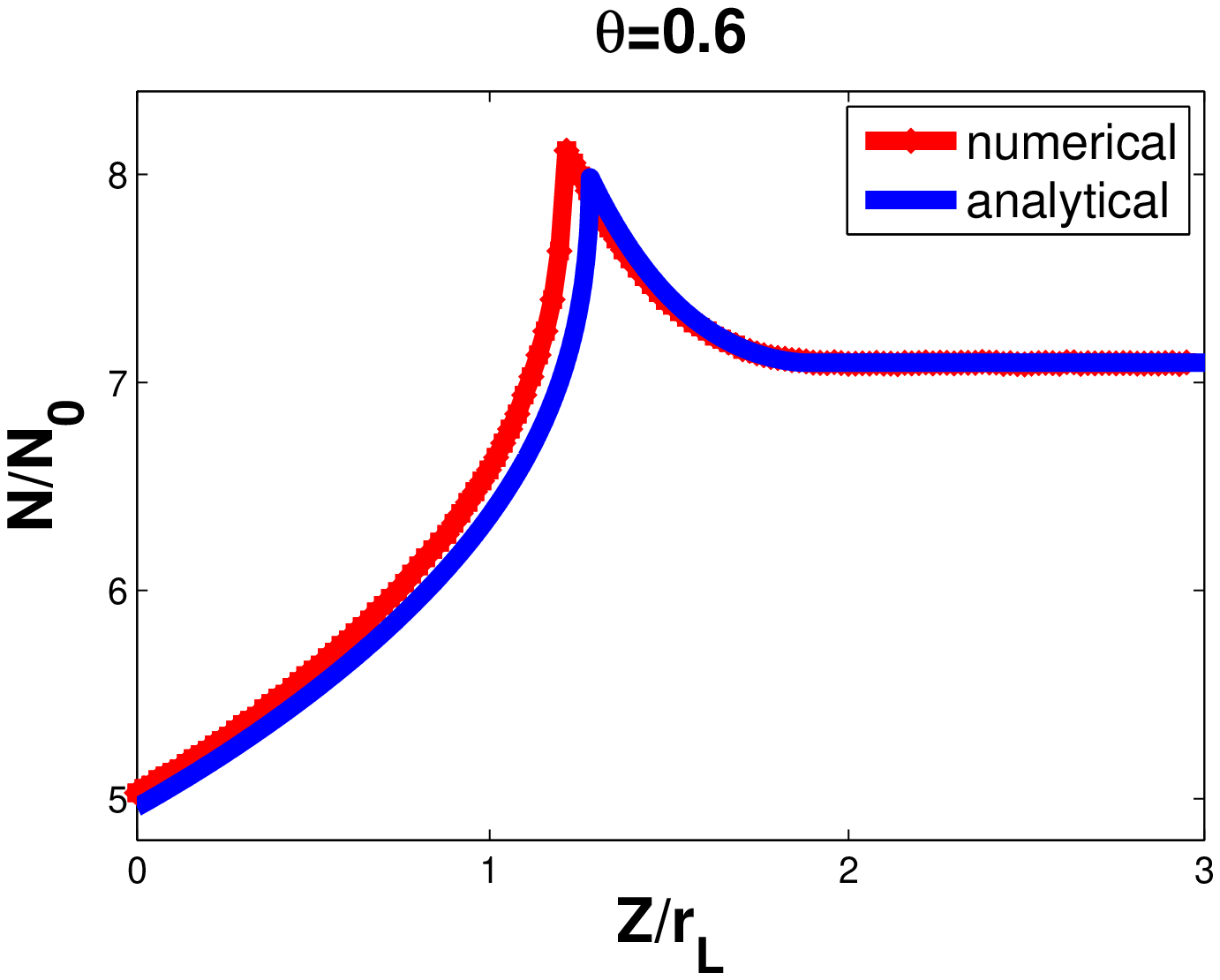}} f) \\
\end{minipage}
\caption{The distributions of the electric current (a,c,e) and number density (b,d,f) of the TCS for the $\theta_0=0.2,0.4,0.6$. The red line is result of numerical simulation and the blue line is the result of the analytic solution eq. (\ref{eq18}).}
\label{fig3}
\end{figure}

\begin{figure}
\center{\includegraphics[width=0.49\linewidth]{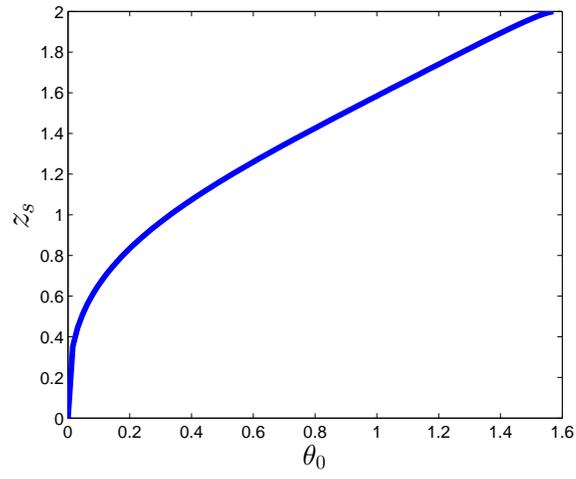}}
\caption{The scaling parameter $z_S$ as a function of $\theta_0$.}
\label{fig4}
\end{figure}

\begin{figure}
\center{\includegraphics[width=0.49\linewidth]{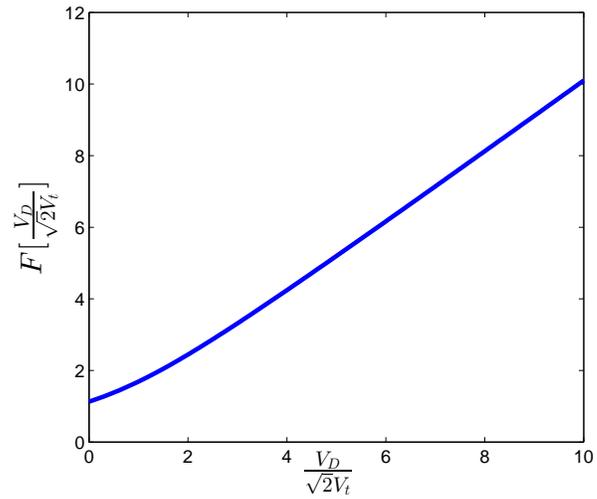}}
\caption{The dependence of coefficient F on the ratio $V_D/\sqrt{2}V_t$.}
\label{fig5}
\end{figure}
\end{document}